\documentclass[11pt]{article}
\usepackage{epsfig}

\newcommand{\be}{\begin{equation}}
\newcommand{\ee}{\end{equation}}

\newcommand{\equ}[1]{Eq.~(\ref{#1})}

\newcommand{\Eqs}[2]{Eqs.~(\ref{#1}) and (\ref{#2})}
\newcommand{\efe}[1]{Ref.\cite{#1}}
\newcommand{\efs}[2]{Refs.\cite{#1,#2}}

\begin{document}
%%%%%%%%%%%%%%%%%%%%%%%%%%%%%%% titlepage %%%%%%%%%%%%%%%%%%%%%%%%%%%%%%%%%%%%
\begin{titlepage}
\begin{flushright}
        \small
         DFPD-99/TH/02\\
         hep-ph/yymmnn\\
         February 1999
\end{flushright}

\begin{center}
\vspace{1cm}
{\large\bf Constraints on the Higgs 
        Boson Mass from Direct Searches and Precision Measurements }

\vspace{0.5cm}
\renewcommand{\thefootnote}{\fnsymbol{footnote}}
{\bf    G.~D'Agostini$^a$ and  G.~Degrassi$^b$ }
\setcounter{footnote}{0}
\vspace{.8cm}

{\it
   $^a$ Dipartimento di Fisica, Universit{\`a} di Roma ``La Sapienza'', \\
     Sezione INFN di Roma 1, P.le A.~Moro 2, I-00189 Rome,    Italy\\
\vspace{2mm}
        $^b$ Dipartimento di Fisica, Universit{\`a}
                  di Padova, Sezione INFN di Padova,\\ 
                  Via F.~Marzolo 8 , I-35131 Padua, Italy\\
\vspace{1cm} }

{\large\bf Abstract}

\vspace{.5cm} 
\end{center}
We combine, within the framework of the Standard Model,
the results of Higgs search experiments with the information 
coming from accurate theoretical calculation and precision measurements to
provide a probability density function for the Higgs mass, 
from which all numbers of interest can be 
derived. The expected value  is
170 GeV, with an expectation uncertainty, quantified 
by the standard deviation of the distribution, of about 80 GeV.
The median of the distribution is 150 GeV, while 75 \%
of the probability is concentrated in the region $M_H \leq 200$ GeV.
The 95\,\% probability upper limit comes out to be around 300 GeV. 
\noindent

\end{titlepage}
%%%%%%%%%%%%%%%%%%%%%%%%%%%%%%%%%%%%%%%%%%%%%%%%%%%%%%%%%%%%%%%%%%%%%%%%%%%%%
% from the dvips manual: put a background `DRAFT' on the page
%  \special{!userdict begin 
%  /bop-hook{gsave 200 30 translate 65 rotate
%           /Times-Roman findfont 216 scalefont setfont
%           0 0 moveto 0.95 setgray (DRAFT) show grestore}def end}
%
%%%%%%%%%%%%%%%%%%%%%%%%%%%%%%%%%%%%%%%%%%%%%%%%%%%%%%%%%%%%%%%%%%%%%%%%%%%%%
\setcounter{page}{2}

\section{Introduction}
Presently, one of the main interests in High Energy physics is
the search for evidence of the Higgs boson and the determination
of its mass. Although all  direct searches have
been unsuccessful till now, 
the self consistency of the Standard Model (SM) in the 
electroweak sector\cite{AB}   
makes physicists highly
confident about the hypothesis 
that the Higgs boson exists, and, most likely,
it has effective properties close to those 
expected from the minimal Standard Model. 

In this paper we use the accurate theoretical predictions for 
the effective mixing parameter, 
$\sin^2{\theta^{lept}_{eff}}\equiv s^2_{eff}$ 
\cite{DGS}, and the $W$ boson mass, $M_W$ \cite{DGV},
together with available experimental information, including also the
results of direct search experiments carried out at LEP, to infer the value 
of Higgs mass, $M_H$. Clearly, the unavoidable status of uncertainty on 
the value of each of the experimental parameters, as well as on the 
accuracy of the 
calculations, allows only a probabilistic inference to be made.  
As a result  we provide a  probability density function (p.d.f.)
for the mass of the Higgs boson
$$
f(m_H\,| \mbox{``data'',``SM''} ) \equiv f(m_H\,|\,\mbox{\it dir.}\, 
 \& \, \mbox{\it ind.}) 
$$
conditioned by the experimental data from direct searches ({\it dir.})
and precision  measurements ({\it ind.})
under the assumption of validity of the SM. 
From this function we
make a  set of probabilistic statements about
$M_H$, and summarize the result in terms of convenient
and  conventional numbers (expected value, standard deviation, mode, median,
etc.). 

The paper is structured in the following way: in the next section 
we recall the theoretical formulae used in the analysis; 
section \ref{anmeth}  is devoted to a detailed description of the 
inferential method used; all the input quantities which enter the analysis
are introduced and commented 
on section \ref{sec:input}.  Section \ref{sec:res_dir} deals with the
determination of $M_W$ and $M_H$ using only indirect information.
Section \ref{sec:combinazione} presents the main result of the paper,
namely $f(m_H\,|\,\mbox{\it dir.}\,  \& \, \mbox{\it ind.}) $.
Finally  we draw some conclusions.
\section{SM formulae relating the Higgs mass to the experimental 
observables}  
The most convenient way to approach the problem is to make use 
of the simple parameterization proposed in 
\efe{Degrassi}, in which  the relations among the 
observables mostly sensible to the Higgs boson mass
are summarized in two formulae:
\begin{eqnarray}
s^2_{eff} &=& (s^2_{eff})_\circ + c_1A_1+c_2A_2-c_3A_3+c_4A_4\,, 
\label{eq:DG3}\\
M_W       &=& M_W^\circ-d_1A_1-d_5A_1^2-d_2A_2+d_3A_3-d_4A_4\,. 
\label{eq:DG4}
\end{eqnarray} 
$(s^2_{eff})_\circ$, $M_W^\circ$,
$c_i$ and $d_i$ are theoretical quantities and $A_i$ are related to 
experimental observables, namely 
$A_1\equiv \ln(M_H/100\,\mbox{GeV})$,
$A_2\equiv\left[(\Delta\alpha)_h/0.0280-1\right]$, 
$A_3\equiv\left[(M_t/175\,\mbox{GeV})^2-1\right]$
and 
$A_4\equiv\left[(\alpha_s(M_Z)/0.118-1\right]$, where $M_t$ is 
the top quark
mass, $\alpha_s(M_Z)$ is the
strong coupling constant and $(\Delta\alpha)_h$ is the 
five-flavor hadronic
contribution to the QED vacuum polarization at $q^2=M_Z^2$. 
The two theoretical quantities 
$(s^2_{eff})_\circ$ and $M_W^\circ$ are the analogues 
of the experimental ones, but obtained by the 
theory at the reference point
$(\Delta\alpha)_h=0.0280$, 
$M_t=175$\,GeV, and $\alpha_s(M_Z)=0.118$. 

\Eqs{eq:DG3}{eq:DG4} are simple analytical formulae that reproduce
to very good accuracy the results  of \efs{DGS}{DGV}. In these papers
the incorporation of the $O(g^4 M_t^2/M_W^2)$ corrections into
the calculation of the effective electroweak mixing angle and $M_W$
is presented. These new contributions are now implemented in the fitting 
codes used by the LEP Electroweak Working Group (EWWG). Their analysis 
shows that the effect of these new corrections is to 
lower the fitted value for  the Higgs mass 
by about $ 30$ GeV and to cause
a sizable decrease in the ambiguity
related to the scheme dependence \cite{EWWG}. Tables 1 and 2 of 
\efe{Degrassi}
report the values of the various theoretical quantities entering
in \Eqs{eq:DG3}{eq:DG4}\ for three different renormalization
schemes. Using these values, \equ{eq:DG3} approximates  
the detailed calculations of \efs{DGS}{DGV} for 
\mbox{75 $\leq M_H \leq 350$} GeV, with the other parameters
in the ranges around the reference values $\delta M_t = \pm 5.5$ GeV, 
$\delta (\Delta \alpha)_h =  \pm 0.00065$ and $\delta \alpha_s = \pm 0.05$,
with average absolute deviations of 
$\approx 4\times 10^{-6}$ and maximum absolute deviations of
$(1.1-1.3)\times 10^{-5}$ depending on the scheme. 
Similarly \equ{eq:DG4}\ shows
average absolute deviations of approximately 0.2 MeV and maximum absolute 
deviations of $(0.8-0.9)$ MeV. Outside the above range, the deviations 
increase reaching $(2.6-2.8)\times 10^{-5}$ and $(3.1-3.3)$ MeV at 
$m_H=600$ GeV.
It is clear that augmenting \Eqs{eq:DG3}{eq:DG4}\ with 
higher powers of the $A_i$ quantities one can reach a better agreement 
in a wider range of Higgs mass values. However, once ``a posteriori''
it is verified that indeed $f(m_H)$ is concentrated in the region 
of validity of \Eqs{eq:DG3}{eq:DG4} then the use of more complicated
parameterization is not really necessary. 
 
\Eqs{eq:DG3}{eq:DG4}\  can be used for two different
purposes. The first is to
solve them with respect to  $A_1$ and $M_W$, getting 
simultaneously  $M_H$ and $M_W$. This value of $M_W$ can be compared
to the experimental measurements in order to check the 
consistency of the theory. Once this check is performed, 
both \Eqs{eq:DG3}{eq:DG4}\  can be used 
to infer $M_H$, although the two determinations
are not independent, due to common terms in the two formulae. 
The combination of these two results, taking into account the correlations,
provides a joint distribution that, further constraint by the 
direct search results,  will give us the final 
 $f(m_H\,|\,\mbox{``data'',``SM''})$. 
\section{Analysis method}\label{anmeth}
\subsection*{Probabilistic approach to infer $M_H$ and $M_W$}
The quantities entering  \Eqs{eq:DG3}{eq:DG4}\  
(hereafter called ``input quantities'') are not known exactly, 
and this makes the result uncertain too. 
It is natural to handle this 
uncertainty by probability. 
The numerical value of the input quantities, which here will be  
generically indicated\footnote{Notice 
that, following the practice of probability
theory, we indicate with capital letters the name of the variable
and with small letters the values they may assume.}
by $X_i$, is interpreted as an 
{\it uncertain number} (also called 
``random variable''). This means that each of them 
could assume an infinite number of possibilities, each 
characterized by a number $f(x_i)$, such that $f(x_i)dx_i$ gives
the (infinitesimal) probability that the ``true value''
$X_i$ is in the interval $dx_i$ 
around $x_i$. 
In this framework the extraction of $M_H$ from \equ{eq:DG3} 
gives a solution that depends on the  uncertain values $X_i$,
\begin{equation}
M_H = M_H(X_1, X_2, \ldots, X_n)\,,\label{eq:M_H} 
\label{eq:M_Hgeneric}
\end{equation}
and therefore we need to evaluate the p.d.f. of a function
of random variables. 
The most general way to describe the uncertainty
about the value of quantities $X_i$ is given by the joint distribution 
$f(x_1, x_2, \ldots, x_n)$. 
Then, a straightforward application of the probability
calculus leads to\cite{dagocern}: 
\begin{equation}
f(m_H) = \int f(x_1, x_2, \ldots, x_n)
\cdot\delta\left(m_H-M_H(x_1, x_2, \ldots, x_n)\right)
\, dx_1dx_2\cdots dx_n\label{eq:f(m_H)}\,,
\label{eq:intdelta}
\end{equation}
where the integral is extended over the hypervolume 
in which  $X_i$ are defined. The l.h.s.\ of \equ{eq:intdelta}\ actually
stands for $f(m_H\,|\,\mbox{{\it ind.}, ``\equ{eq:DG3}''})$.
\equ{eq:intdelta} has a simple 
intuitive interpretation\footnote{An alternative way of 
interpreting 
(\ref{eq:intdelta}) is to think to a Monte Carlo simulation, where all 
the input quantities enter with their distributions, and 
correlations are properly taken into account. The histogram of 
$M_H$ calculated from (\ref{eq:M_Hgeneric}) will ``tend'' to 
$f(m_H)$ for a large number of generated events.}:
 the (infinitesimal) probability 
element $f(m_H)\,dm_H$ depends on ``how many'' elements 
$dx_1dx_2\cdots dx_n$
contribute to it, each weighted by the p.d.f. calculated in 
the point $\{x_1, x_2, \ldots, x_n\}$.

The solution of \equ{eq:intdelta} is very complicated, however we can
perform a series of
approximations and use of  central limit
theorem to get the final p.d.f. without actually making explicit 
use of Eq. (\ref{eq:intdelta}) and without reducing 
the accuracy of the inference.  
We would like to list the steps needed to determine
$f(m_H\,|\,\mbox{\it ind.},\,\mbox{``\equ{eq:DG3}''})$. 
\begin{itemize}
\item
First, with a great degree of approximation, the quantities entering 
\equ{eq:DG3} are independent, or at least 
this condition is satisfied for the quantities
from which the uncertainty on $M_H$ mostly depends. Actually,
the theoretical parameters entering \Eqs{eq:DG3}{eq:DG4}\ contain the
same information evaluated in
different  renormalization schemes, and, therefore, they could 
all be correlated. A more careful procedure for handling their 
uncertainty could be considered. This issue
will be discussed at the end of this paragraph, 
and the numerical outcomes of the two methods used will be compared
when discussing the results.  
\item
Second, we make use of the central limit theorem,
which makes
the probability distribution
of a linear combination of random quantities 
under well known conditions Gaussian.
The importance of this theorem is that we only have 
to make sure that the terms dominating the overall uncertainty
are practically Gaussian. As far as the other terms are
concerned, 
the exact form of the individual distributions doesn't even matter,
since  only 
expected value and variance are relevant.
\item 
The consequences of the central limit theorem can be 
extended to the variables which do not enter linearly,
if their dependence  can be linearized with a 
reasonable degree of approximation in a range of several 
standard deviations around their expected value. This amounts  
to requiring these variables to have a sufficiently small 
{\it variation coefficient} (the ``relative uncertainty'' of 
the physicists' lexicon).
\item 
Applying this analysis to our case, 
we see that the solution of \equ{eq:DG3} in terms
of $m_H$  is strongly not linear. Therefore  
$A_1$ is the natural quantity
with which to express the result at an intermediate stage, 
being 
\begin{equation}
A_1 \sim {\cal N}(\mbox{E}[A_1], \sigma(A_1))\,,
\end{equation}
where the last notation is a shorthand for a normal distribution
of expected value $\mbox{E}[A_1]$ and standard deviation 
$\sigma(A_1)$, calculated as
\begin{eqnarray}
\mbox{E}[A_1] &=& A_1(\mbox{E}[X_1], \mbox{E}[X_2], \ldots, 
               \mbox{E}[X_n]) \\
\sigma^2(A_1) &=&  
\sum_i\left(\frac{\partial A_1}{\partial X_i}\right)^2\sigma^2(X_i)\,
\end{eqnarray}
with the derivates evaluated at the expected values.
\item
Finally, the exact form of $f(m_H)$ can be obtained from $f(a_1)$, 
making use of standard probability calculus, 
e.g. using \equ{eq:intdelta}.
\end{itemize}
A similar strategy can be  used to get the parameters of the Gaussian
which describes the knowledge of $M_W$. In this case the linearization 
hypothesis is already reasonable  for  $M_W$ itself
and the 
resulting $f(m_W)$ is therefore normal with a  high
degree of approximation. 

The procedure outlined above does not take into account  possible correlations
among the theoretical parameters of \Eqs{eq:DG3}{eq:DG4}. 
Estimating  the correlation coefficients from the sample
provided by  Tables 1 and 2 of \efe{Degrassi} would 
be a  rough and complicated  procedure.
In fact the covariance matrix can only
take into account linear correlations, whether, in general, these effects 
could be more subtle. 
A more elegant and general way to handle this information is, then,
to consider different inferences,
each conditioned by a given set of parameters, labelled by $R_i$.
This can be 
applied at any stage of the analysis, although it is in practice more
convenient to apply it at the level of the inference on $A_1$. 
For each renormalization scheme $R_i$ we have then:
\begin{equation}
A_1 |_{R_i}\sim {\cal N}(\mbox{E}[A_1\,|\,R_i], \sigma(A_1\,|\,R_i))\,.
\end{equation}   
The p.d.f. of $A_1$, ``integrated'' over the possible schemes, is then
\begin{equation}
f(a_1)=\sum_i f(a_1\,|\,R_i)\cdot f(R_i)\,,
\label{eq:fA1av}
\end{equation}
where $f(R_i)$ is the probability assigned to each scheme. 
The calculation of expectation value and variance
is straightforward. When there is indifference with respect to
the renormalization schemes (i.e. $f(R_i)=1/3$ $\forall i$) we get 
\begin{eqnarray}
\mbox{E}[A_1] &=& \frac{1}{3}\sum_i \mbox{E}[A_1\,|\,R_i] \\
\sigma^2(A_1) &=& \frac{1}{3}\sum_i \sigma^2(A_1\,|\,R_i) +
                  \frac{1}{3}\sum_i\mbox{E}^2[A_1\,|\,R_i]-\mbox{E}^2[A_1]\\
              &=& \frac{1}{3}\sum_i \sigma^2(A_1\,|\,R_i) + 
                  \sigma^2_E\,,
\end{eqnarray}
where $\sigma_E$ indicates the standard deviation calculated from the 
dispersion of the 
expected values. 
The p.d.f. (\ref{eq:fA1av}) is in general 
not Gaussian, since it comes from a linear combination
of p.d.f.'s, and not from a linear combination of variables
(i.e. the central limit theorem does not apply).
Nevertheless, in our case the Gaussian approximation will be valid, as 
will be discussed below. 
\subsection*{Double inference on $M_H$ and combination of the results}
The method  described in the previous section 
is applied to each of the 
\Eqs{eq:DG3}{eq:DG4}, obtaining two inferences on $A_1$,   
the first (indicated by $A_1^s$) depending on the effective electroweak
mixing parameter and the second ($A_1^w$)
on the $W$ mass. 
The second equation
leads to two solutions and the largest value has been considered, 
because of the agreement with the $A_1^s$ and also because the smaller
solution leads to a mass well below the range firmly excluded by 
present observations.

The two uncertain values $A_1^s$ and $A_1^w$ are 
not independent, due to the fact that some of the 
input quantities appear in both 
relations. This means that we have to consider the joint distribution 
$f(a_1^s, a_1^w)$. Because each variable is individually Gaussian, 
the joint distribution is described by a two-dimensional normal, 
with a correlation coefficient  
$\rho(A_1^s,A_1^w)$ 
calculated from the covariance between $A_1^s$ and $A_1^w$: 
\begin{eqnarray}
\rho(A_1^s, A_1^w)  &=& 
\frac{\mbox{Cov}(A_1^s,A_1^w)}{\sigma(A_1^s)\cdot\sigma(A_1^w)} 
\nonumber
\end{eqnarray}
Again using  linearization around the expected values, one finds 
easily that the covariance is given by
\begin{eqnarray}
\mbox{Cov}(A_1^s,A_1^w) &=& \sum_i\frac{\partial A_1^s}{\partial X_i}\cdot
               \frac{\partial A_1^w}{\partial X_i}\cdot\sigma^2(X_i) 
\nonumber \\
           &=& \sum_i  
%\left(\mbox{sign}\left[\frac{\partial A_1^s}{\partial X_i}\right]\cdot 
%\sigma_i(A_1^s)\right) \cdot
%\left(\mbox{sign}\left[\frac{\partial A_1^w}{\partial X_i}\right]\cdot 
%\sigma_i(A_1^w)\right)\,, \label{eq:cov}
\left(\frac{\partial A_1^s}{\partial X_i}\cdot 
\sigma(X_i)\right) \cdot
\left(\frac{\partial A_1^w}{\partial X_i}\cdot 
\sigma(X_i)\right)\,, \label{eq:cov}
\end{eqnarray} 
where this last formulation is very convenient for practical purposes,
as we will see below.
Eq.(\ref{eq:cov}) does not take into account the correlations among the 
various theoretical coefficients. 
However, numerically they are completely negligible
with respect to the ones due to $A_2,\,A_3$ and $A_4$ and therefore the 
use of
\equ{eq:cov}\ is well justified. 
%
%although not usual in literature (see \cite{primer}). 

The presence of the correlation term prevents the two results
from being combined
with the usual formula of the average weighted with the inverse 
of the variance. There are several possibilities 
for  taking correlation into account, 
either working directly with p.d.f.'s, or assuming that 
the final result is also normally distributed and evaluating the two
parameters of the distribution. Obviously the conclusions will not depend
on the procedure if the normality 
assumption is correct, as it is in this case. 
The way that seems to us the most intuitive relies on the fact that 
$\rho(A_1^s, A_1^w)$ is positive, as we will see, and thus the
correlation between the two results is equivalent to that introduced by 
an uncertainty on a common offset (see, e.g., \cite{matrix}).
Therefore  the variances
of $A_1^s$ and $A_1^w$ 
may be considered as being 
formed of two parts: one of these parts, indicated 
by $\sigma_c^2$, is common to both variances; while the other
is individual. 
The common part is given by covariance, i.e.
$\sigma_c^2=\mbox{Cov}(A_1^s,A_1^w)$. 
The individual contribution to  each variance is then evaluated
subtracting $\sigma_c^2$. 
This procedure allows the expected value $\mbox{E}[A_1]$ 
to be evaluated as the 
average of $\mbox{E}[A_1^s]$ and 
$\mbox{E}[A_1^w]$, weighted with the inverse of the individual 
variance. The variance $\sigma^2(A_1)$ will be, finally,  
the sum of the 
``variance of the weighted average'', plus the common 
variance.\footnote{An alternative way, which still avoids  working 
with p.d.f.'s, 
is described in \cite{Cowen}. The two procedures yield identical results.}  
We have, then:
\begin{eqnarray*}
\mbox{E}[A_1] &=& \left(\frac{\mbox{E}[A_1^s]}{\sigma^2(A_1^s)-\sigma_c^2}+
             \frac{\mbox{E}[A_1^w]}{\sigma^2(A_1^w)-\sigma_c^2} \right)
         \left(\frac{1}{\sigma^2(A_1^s)-\sigma_c^2}+
             \frac{1}{\sigma^2(A_1^w)-\sigma_c^2}\right)^{-1} \\
\sigma^2(A_1) &=& \sigma_c^2 + 
             \left(\frac{1}{\sigma^2(A_1^s)-\sigma_c^2}+
             \frac{1}{\sigma^2(A_1^w)-\sigma_c^2}\right)^{-1}     
\end{eqnarray*}
Finally, the Gaussian result on $A_1$ is transformed into the p.d.f. 
of $M_H$ using probability 
calculus. For example, one can make use of \equ{eq:intdelta},
and the result is straightforward, remembering that 
$\delta(m_H-100\,\exp a_1)=m_H^{-1}\delta(a_1-\ln(m_H/100))$.
Then it is possible
to evaluate expected value, standard deviation, mode ($\hat{M}_H$)
 and median ($M^{50}_H$) of $M_H$ (see, e.g., 
\cite{JK} for the properties of the so called 
{\it lognormal} distribution). 
The results (expressed in GeV) are:
\begin{eqnarray}
f(m_H\,|\,\mbox{\it ind.}) &=& \frac{1}{\sqrt{2\,\pi}\,\sigma(A_1)}        
\frac{1}{m_H}
         \exp{\left[-\frac{\left(\ln{(m_H/100)}-\mbox{E}[A_1]\right)^2}
                          {2\,\sigma^2(A_1)}\right]}
\label{eq:m_h_da_a1} \\
\mbox{E}[M_H] &=& 
100\,\exp\left[\mbox{E}[A_1]+\frac{1}{2}\,\sigma^2(A_1)\right] 
\label{eq:Em}\\
\sigma(M_H) &=& 100\,\left(
\exp{\left[2\,\mbox{E}[A_1]+2\,\sigma^2(A_1)\right]}-
\exp{\left[2\,\mbox{E}[A_1]+\sigma^2(A_1)\right]}\right)^{\frac{1}{2}}
\hspace{0.6cm} 
\label{eq:sigmam}\\
\hat{M}_H &=& 100\,\exp{\left[\mbox{E}[A_1]-\sigma^2(A_1)\right]}
\label{eq:modem}\\
M^{50}_H &=& 100\,\exp{\left[\mbox{E}[A_1]\right]}
\label{eq:medianm}
\end{eqnarray}
Notice that the value of these position and dispersion 
parameters of $f(m_H)$ is, in general, 
not simply the back transformation 
of those of $f(a_1)$.
\subsection*{Including the constraint from  direct search} \label{sec:analysis}
The knowledge about the value of $M_H$ is  modified further by the 
non-observation of the Higgs boson up to the highest LEP energies.  
To understand how $f(m_H\,|\, \mbox{\it ind.})$ 
changes when it is further conditioned 
by the negative direct experimental result, let us consider 
a search for Higgs production in association with a particle of negligible 
width in an ideal situation (``infinite'' luminosity,
perfect efficiency,  no background) whose outcome was
  no candidate. Consequently  all mass 
values below a  sharp kinematical 
limit $M_K$ are excluded. 
This implies that: a) $f(m_H)$ must vanish below $M_K$  
(otherwise one would have observed the particle);
b) above $M_K$ the relative probabilities cannot change, because 
there is no sensitivity in this region, and then the experimental
results cannot 
give information over there. 
For example, if $M_K$
is 90 GeV, then $f(200\, \mbox{GeV})/f(100\, \mbox{GeV})$ 
must remain constant before and after 
the new piece of information is included. 
In this ideal case we have then
\begin{equation}
f(m_H\,|\,\mbox{\it dir.} \,\&\,\mbox{\it ind.}) = 
\left\{\begin{array}{ll} 
0   & m_H < M_K  \\
 \frac{f(m_H\,|\,\mbox{\it ind.})}
    {\int_{M_K}^\infty f(m_H\,|\,\mbox{\it ind.})\, dm_H} & m_H \ge M_K\,, 
\end{array}\right.
\label{eq:taglio_secco} 
\end{equation}   
where the integral at denominator is just a normalization coefficient. 

More formally, this result can be obtained making explicit use
of the Bayes' theorem. Applied to our problem, the theorem
can be expressed as follows (apart from a
 normalization constant):
\begin{equation}
f(m_H\,|\, \mbox{\it dir.} \,\&\,\mbox{\it ind.})
 \propto f(\mbox{\it dir.}\,|\,m_H)\cdot  f(m_H\,|\,\mbox{\it ind.})\,,
\label{eq:Bayes}
\end{equation} 
where $f(dir\,|\,m_H)$ is the so called  likelihood, 
which has the role of updating the p.d.f.~once the new piece of
information is included in the inference.
In the idealized example  we are considering now, 
$f(dir\,|\,m_H)$ can be expressed in terms of
the probability of  observing zero candidates in an experiment sensitive 
up to a $M_K$ mass for a given value $m_H$, or
\begin{equation}
f(\mbox{\it dir.} \,|\, m_H) = f(\mbox{``zero cand.''}\,|\,m_H) = 
\left\{\begin{array}{ll} 
0   & m_H < M_K \\
      1 & m_H \ge M_K \, . 
\end{array}\right.
\label{eq:lik_taglio_secco} 
\end{equation}   
In fact, we would expect an ``infinite'' number of events 
if $M_H$ were below the kinematical limit.
Therefore the probability of observing nothing should be zero. 
Instead, for $M_H$ above $M_K$,
the condition of vanishing production cross section and no background
can only yield no candidates. 

\begin{figure}
\begin{center}
\epsfig{file=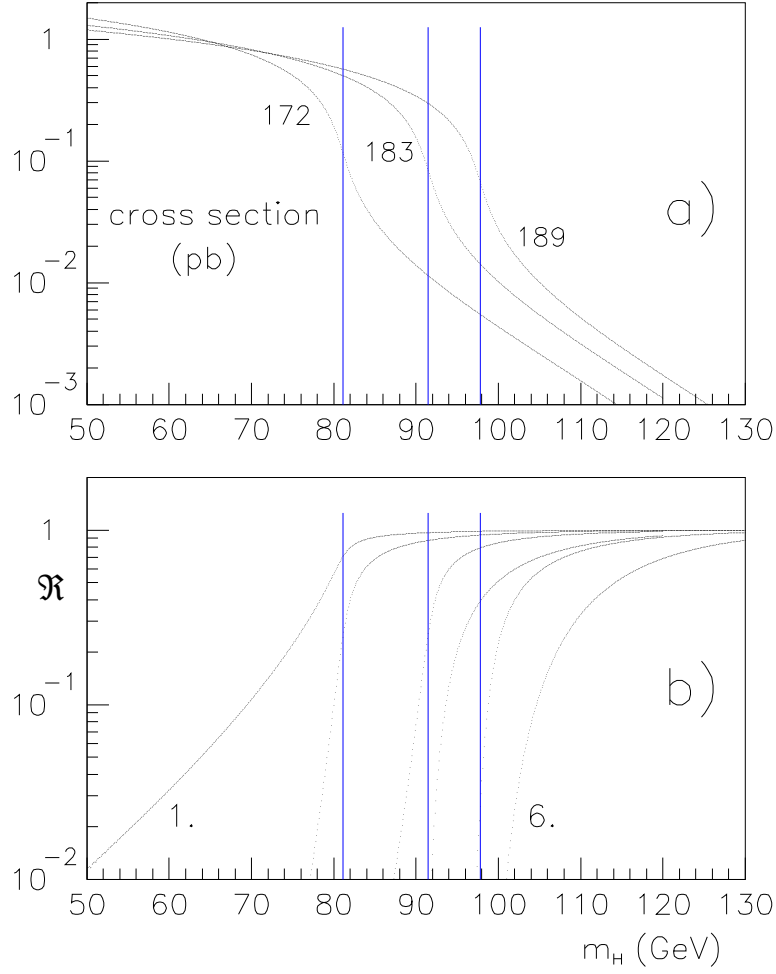,clip=,width=\linewidth}
\end{center}
\caption{\sf  a) cross-section $e^+ e^- \rightarrow H+Z^\circ$ as a 
function of
$m_H$ for $\sqrt{s} = 172,\, 183,\, 189$ GeV. The vertical lines are at
$m_H = \sqrt{s} - M_Z$. b) ${\cal R}$ vs.~$m_H$ for
$n_{obs} = \lambda_B= 0$ for $\sqrt{s} = 172,\, 183,\, 189$ GeV, with 
$\epsilon = 30 \%$ and ${\cal L} = 10,\, 55,\, 180\: {\rm pb}^{-1}$ per 
experiment, respectively. Odd lines are for a single experiment, even ones
represent the combination of four.}
\label{fig:sigma_lik}
\end{figure}

In real life situations 
the transition between  values which are impossible to 
those which are possible is not so sharp. 
Because of physical reasons (such as threshold effects
and background)
and experimental reasons (such as luminosity (${\cal L}$) 
and efficiency ($\epsilon$)) we cannot be  really 
sure about excluding values just below $M_K$, nevertheless
very small values of the mass are ruled out. 
In the case of Higgs production  at LEP the dominant mode 
is the Bjorken process $ e^+  e^- \rightarrow H + Z^\circ$. 
This reaction does not have  a sharp kinematical
limit at $\sqrt{s}-M_Z$ (minus a negligible kinetic energy), due to the 
large total width of the $Z^\circ$. The effective 
kinematical limit ($M_{K_{eff}}$) 
depends, then, on the available integrated 
luminosity and  could reach up 
to the order of $\approx\sqrt{s}-M_Z+{\cal O}(10\,\mbox{GeV})$ 
for very high luminosity. This is clear from figure 
\ref{fig:sigma_lik}a, where the cross section 
$e^+e^-\rightarrow H + Z^\circ$, with the $Z$ decaying in all possible 
channels, is plotted as a function
of the Higgs boson mass for 172, 183 and 189 GeV c.m. energy,  
with the vertical lines showing $\sqrt{s}-M_Z$ for the three cases.  

In this  non ideal situation we expect the step function of  
 \equ{eq:lik_taglio_secco}   to
be replaced by a smooth curve which goes to zero
for low masses  and to 1 for $M_H\rightarrow M_{K_{eff}}$, 
where the experimental sensitivity is lost. 
In this case {\it dir.} in \equ{eq:Bayes} stands, in principle, for  
all possible experimental observables and the function 
$f(\mbox{\it dir.}\,|\,m_H)$
should be provided by the experiments. However the LEP
collaboration results  on the Higgs mass searches are usually presented
in terms of confidence level $(C.L.)$ \cite{LEP4}.
As discussed in section \ref{sec:combinazione} this  quantity  
does not have a simple  connection
to $f(\mbox{\it dir.}\,|\,m_H)$. Given this situation we decided to model
the likelihood in a way which seems compatible with the physics case. 

First, all possible experimental observables can, 
in practice,  be replaced by suitable combinations 
which depend on the Higgs mass. The simplest 
of these possible ``summaries'' of the data is the number 
of observed candidate events, which we will indicate by $n_{obs}$.  
The number of candidate 
events expected to be observed, on the other hand, is  
given by the sum of the Higgs events, indicated by 
$\lambda_S(m_H)$, and the expected number 
of background events, $\lambda_B$, assumed here to be well known
(see e.g.~\efe{dagocern} for the natural extension when 
$\lambda_B$ is uncertain too). 
The mass dependence of the former is due to the
mass dependence of cross section, branching ratio $(b.r.)$ and efficiency, 
and so it depends
on the decay channel  investigated. For simplicity, we discuss the case of
a likelihood obtained considering the total number of observed candidate 
events in a single  channel. This  is given by 
\begin{equation}
f(n_{obs}\,|\,m_H,\lambda_B) = \frac{e^{-(\lambda_S(m_H)+\lambda_B)}
                              \cdot(\lambda_S(m_H)+\lambda_B)^{n_{obs}}}
                              {n_{obs}!}\,,
\label{eq:lik_Poisson}
\end{equation}
since $n_{obs}$ is expected to be described by a Poisson distribution 
with parameter $\lambda = \lambda_S(m_H)+\lambda_B$. 
In order to compare and combine the updating power provided
by each piece of information easily, it is convenient to replace
the likelihood by a function, ${\cal R}$,
that goes to 1 where the experimental sensitivity is 
lost \cite{dago_limits}.
This function can be seen as the counterpart, in the case of a real 
experiment, of the step function of \equ{eq:lik_taglio_secco}. 
Because constant factors do not play any role in the Bayes' theorem
we can divide the likelihood by its value
calculated for very large Higgs mass values, where no signal 
is expected\footnote{In the statistics lexicon this 
function is the {\it Bayes factor} between the generic mass $m_H$ 
and $M_H=\mbox{``}\infty\mbox{''}$.}, 
i.e. $M_H\rightarrow\mbox{``}\infty\mbox{''}$, or
$\lambda_S\rightarrow 0$. Clearly, this operation makes sense only if the 
likelihood is different from 0 for $M_H\rightarrow\mbox{``}\infty\mbox{''}$. 
This condition is satisfied for any $\lambda_B$
in case  $n_{obs}\ne 0$, but when $n_{obs}=0$ only for $\lambda_B=0$.
The case of $n_{obs}\ne0$
with $\lambda_B=0$ leads to a clear discovery, i.e. 
the likelihood will assume positive values only below 
$M_{K_{eff}}$ and there is no need anymore to build 
the ${\cal R}$ function with the desired asymptotic properties.

${\cal R}$, as a mathematical function of $m_H$, 
with $n_{obs}$ and $\lambda_B$ acting as parameters,
can be seen as a kind of shape distortion function of the 
p.d.f.~introduced by the new data. As long as ${\cal R}(m_H)$ 
is 1, the shape (and therefore the relative probabilities in that region)
remains unchanged, while in the limit  ${\cal R}(m_H)\rightarrow 0$
the p.d.f. vanishes. One should notice that 
 ${\cal R}(m_H)$ can also assume values larger than 1 in the region 
of sensitivity, corresponding to a number of observed candidate
events larger than the expected background. 
In this case the p.d.f.~will be  stretched 
below the effective kinematical limit and this might even
prompt a claim for a discovery if ${\cal R}$ becomes sufficiently large 
for the probability of $M_H$ in that region to get 
very close to 1. 

\begin{figure}[t]
\begin{center} 
\epsfig{file=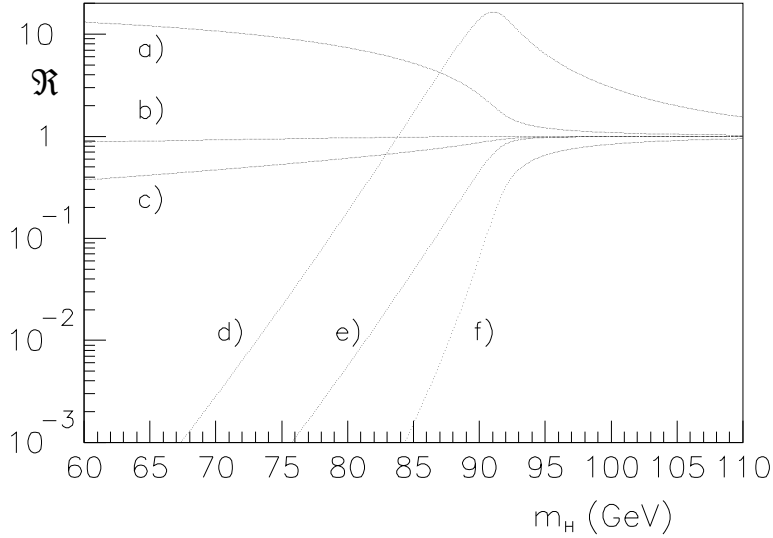,clip=,width=\linewidth}
\end{center}
\caption{\sf ${\cal R}$ vs.~$m_H$ for two search
channels at $\sqrt{s} = 183$ GeV with ${\cal L} = 55\,{\rm pb}^{-1}$.
The lines a--c represent a channel with
$b.r. \times \epsilon = 1.7 \%$ while \mbox{d--f} correspond to a search
with $b.r. \times \epsilon = 32 \%$. The cases considered are:
$n_{obs} = 2,\,\lambda_B= 0.2 \: \mbox{(a,\,d)};\: 
 n_{obs} = 3,\,\lambda_B= 3 \: \mbox{(b,\,e)};\: 
 n_{obs} = 0,\,\lambda_B= 1.5 \: \mbox{(c,\,f)}$.  }
\label{fig:lik_six}
\end{figure}

Applying this formalism to our case and in the 
realistic situation of non vanishing expected background  we get
\begin{equation}
{\cal R}(m_H;n_{obs},\lambda_B) = \frac{e^{-(\lambda_S(m_H))}
                                \cdot(\lambda_S(m_H)+\lambda_B)^{n_{obs}}}
                          {\lambda_B^{n_{obs}}}\,,~~(\lambda_B\neq 0,
\: \mbox{if}~~n_{obs} \neq 0).
\label{eq:bur}
\end{equation}
Instead, when $\lambda_B=0$ and $n_{obs} =0$ one  can take the limit of the 
above formula obtaining ${\cal R}(m_H)=e^{-\lambda_S(m_H)}$.
Examples of this function are shown in figure \ref{fig:sigma_lik}b)
in case of no events candidates and zero background ($n_{obs}=\lambda_B=0$),
for $\sqrt{s}=  172,\, 183$ and 189 GeV,
considering a single LEP experiment (odd numbers) and 
the combination of all experiments (even numbers). The calculations
have been done assuming a nominal integrated luminosity per experiment
of 10, 55 and 180 pb$^{-1}$ for the three energies, and an
average and constant detection efficiency of 30\,\%.
Fig.\ref{fig:lik_six}, instead,  illustrates six different possible scenarios.
In the figure we consider a search  at
$\sqrt{s} = 183$ GeV with ${\cal L} =55 \,{\rm pb}^{-1}$ by a single experiment
that looks for two Higgs decay channels with different branching ratio. 
For each channel we plot 3 different situations of expected 
background and observed number of events. 

As already pointed out, $f(\mbox{\it dir.}\,|\,m_H)$,
and hence ${\cal R}(m_H;\mbox{\it dir.})$,  
might be more complicated
than the simplified likelihood used here, and it 
can  only be provided 
by experiments. Either of the above functions would be the most unbiased 
way of reporting
the experimental result and it would allow 
several pieces of experimental information
to be easily combined.
In fact, when   individual experiments 
or  decay channels are independent the overall likelihood is simply the 
product of the individual likelihoods and therefore
\begin{equation}
{\cal R}(m_H;\mbox{``all data''})
= \Pi_i {\cal R}_i(m_H;\mbox{``data $i$''})\,,
\label{eq:lik_over}
\end{equation}
and this can be used in \equ{eq:Bayes} to get the 
distribution of $M_H$ which takes into account all 
available data.
\section{Input quantities entering the indirect determination}\label{sec:input}
In this section we discuss the experimental and theoretical inputs 
used in our analysis. 
\subsection*{Hadronic contribution to QED vacuum polarization}
The QED coupling at the $Z$ boson mass scale  plays an important role
in the prediction of $m_H$. This fact has always stimulated a lot of activity
on the exact determination of $(\Delta\alpha)_h$. The most  phenomenological
analyses of it
rely on the use of  all the available experimental data on the
hadron production in $e^+ \,e^-$ annihilation and on perturbative QCD (pQCD)
for the high energy tail ($E \geq 40$ GeV) of the dispersion integral.
The reference value in this approach is \cite{Jeg}
\be
\label{EJalpha}
(\Delta\alpha)_h^{EJ} = 0.02804\pm0.00065~.
\ee

In the recent past the hadronic contribution to the vacuum polarization
has been the subject of several new investigations that advocate the
use of pQCD down to energy scale of the order of 1 GeV \cite{MZ,DH}. 
In this more theory driven path, the various analyses
differ on the energy value 
at which to start applying pQCD and on the amount of
theoretical inputs used to evaluate the experimental data in the regions,
like, for example, the threshold for the charmed mesons, where pQCD is not 
applicable. 
The common characteristic of these works  with respect to the most 
phenomenological ones is to obtain a smaller central value for
$(\Delta\alpha)_h$ with a reduced uncertainty. 
The most stringent result
of these theory oriented analyses 
is \cite{DH}
\be
\label{DHalpha}
(\Delta\alpha)_h^{DH}=0.02763\pm0.00016 
\ee
that we will use in the sequel as reference value for this kind of 
approach. 

At the moment there is no definite argument for choosing one or other
of the two approaches.  
The results are absolutely compatible to each other. However,
the numerical difference between  central values and 
uncertainties is such that it prevents  an
easy estimation of the effect 
of choosing one value instead 
of the other.
For these reasons we decided 
to present our results 
for the values of $(\Delta\alpha)_h$ given by 
$(\Delta\alpha)_h^{EJ}$  and  $(\Delta\alpha)_h^{DH}$ 
separately. 
\subsubsection*{Top quark mass}
The value of the top quark used in our analysis is the combination of the 
experimental direct measurements reported by
CDF \cite{CDF} and D0 \cite{D0}: 
\be
M_t=174.2\pm 4.8~\mbox{GeV}. \label{mtop}
\ee
The $M_t$ value obtained in the global fit of the EWWG \cite{EWWG},
that uses as experimental  value input $ M_t = 173.8 \pm 5.0$ GeV, 
is actually a little bit smaller, i.e. $ M_t = 171.1 \pm 4.9$ GeV. 
The principal cause for this smaller value is connected with 
the remnant of the famous $R_b$ ``anomaly''.
In our analysis we assume the
validity of the SM and therefore we prefer to use the experimental
result of \equ{mtop}.
\subsubsection*{QCD strong coupling constant}
Among the various input quantities, the strong coupling constant at the
$M_Z$ scale is the least important. In fact, QCD effects appear in the
theoretical calculations of $s^2_{eff}$ and $M_W$ only at the two
loop level. We use the world average \cite{PDG}
\be
\alpha_s (M_Z) = 0.119 \pm 0.002
\ee
\subsubsection*{Effective mixing parameter $\sin^2\theta_{eff}^{lept}$}
The effective Weinberg angle is the quantity that has the greatest 
sensitivity to the Higgs. Therefore its precise value
is very important in determining $f(m_H)$. 
There is overall good agreement among all the measurements although
the two most precise ones, 
i.e. $A_{{\rm LR}}$ from SLD ($s^2_{eff} =
0.23109 \pm 0.00029$) and $A_{{\rm FB}}^{0,{\rm b}}$ from LEP ($s^2_{eff} =
0.23225 \pm 0.00038$), are still about two and a half 
standard deviations apart.
However, the continual raising of the SLD value 
during the recent years together
with some reduction of the  $A_{{\rm FB}}^{0,{\rm b}}$ result has 
significantly improved the agreement between the average LEP and SLD
determinations. 
In this situation we do  not see any particular reason 
either for excluding the SLD values or for
attributing to it a smaller weight \cite{Chan}.
Therefore we 
employ in our analysis the combined LEP+SLD
average\cite{EWWG}
\be
\sin^2\theta_{eff}^{lept} = 0.23157 \pm 0.00018 .
\label{seff}
\ee
\subsubsection*{$W$ boson mass}
The present experimental information on $M_W$ comes from the 
invariant mass of  its decay products (LEP and Tevatron),
from the threshold behaviour of the production cross section
(LEP) and from the electroweak coupling constant 
in  neutrino scattering (NuTeV,CCFR). The first two measurements
can be considered a kind of ``direct'' determinations of the mass,
in the sense that they are sensitive directly  to it and 
not to a combination of other parameters of the SM.
The combination of CDF \cite{CDFMw}, D0 \cite{D0Mw} and
LEP values \cite{EWWG} (including also the old UA2 measurement \cite{UA2}),
gives\cite{EWWG}:
\be
M_W^k=80.39\pm 0.06~\mbox{GeV}. \label{eq:M_W_kin}
\ee
The result of the deep inelastic scattering  experiments can be reported using 
the quantity
$\sin^2\theta_w=1-M_W^2/M_Z^2$. In this case $M_W$ can be extracted
in terms of the very 
precise $M_Z$ value plus top quark and Higgs boson mass 
corrections \cite{NuTeV}:
\begin{equation}
\frac{M_W^{dis}}{\mbox{GeV}} = \frac{M_W^\nu}{\mbox{GeV}} 
+ 0.073\,\left(\frac{M_t^2-(175\,\mbox{GeV})^2}
                                  {(100\,\mbox{GeV})^2}\right)
              - 0.025\,\ln(M_H/150\,\mbox{GeV})\,,
\label{eq:M_W_NuTeV}
\end{equation}
where $M_W^\nu$ indicates   the result at
the reference values ($M_t=175\,\mbox{GeV}$ and $M_H=150\,\mbox{GeV})$: 
$$M_W^\nu=80.25\pm 0.11\,\mbox{GeV}\,.$$
In order to make use of all available information we proceed in the 
following way. We evaluate $A_1^w$ from $M_W^k$ and $M_W^{dis}$
separately using  \equ{eq:DG4} (in the $M_W^{dis}$ case
the Higgs and top dependence can be accounted for by redefining 
the theoretical coefficients $d_1$, $d_3$, and $M_W^\circ$).
Once the compatibility of the two results has been established, we are
allowed to combine directly the $M_W$
values weighting them with the inverse of the variance. 
We obtain 
\be
\frac{M_W}{\mbox{GeV}} = 
80.36 \pm 0.05 +0.0023 - 0.0057 A_1 + 0.051 A_3 \label{mwcomb}
\ee 
which is the value employed in the analysis. Again, the Higgs and
top dependence is taken into account by slightly modifying the relevant
coefficients in \equ{eq:DG4}.
\subsection*{Theoretical coefficients}
The various coefficients entering  
\Eqs{eq:DG3}{eq:DG4} are
not known exactly  due to truncation of the perturbative series.
This uncertainty is  usually estimated
comparing the results of different schemes of calculation that 
contain all the available theoretical
information at a given order of accuracy. 
Then the simplest procedure is to evaluate the best value and
standard deviation associated 
to the uncertainty of each of the coefficients from the 
average and standard deviation of the values given in \efe{Degrassi}
(when all renormalization schemes yield the same 
numerical results the standard deviation is that due to the rounding, i.e.
unit of the least significant digit divided by $\sqrt{12}$). 
They are indicated in tables 
\ref{tab:MW}, \ref{tab:Jeg} and \ref{tab:A1sum} and considered independent 
in the uncertainty propagation. Let us comment on the 
meaning and the use of averages and  
standard deviations for the coefficients. Taking as an example 
$c_1$,  we get (in units $10^{-4}$) 
$\mbox{E}[c_1]=5.23$ and 
$\sigma(c_1)=0.04$, obtained by the following
numbers\cite{Degrassi}: 
5.23,  
5.19 and
5.26. This does not imply that necessarily one 
has to believe that 5.23 is really more preferred 
than the others, as a Gaussian distribution centered 
in 5.23 with standard deviation 0.04 would imply. One could imagine  
a uniform distribution ranging between 5.16 and 5.30; or 
a triangular distribution centered in 5.23 and going to
zero at 5.13 and 5.23; or any other distribution having mean 5.23
and sigma 0.04.
The final result, relying on the central
limit theorem, which, for the relative sizes of the standard deviations
of interests ensures a fast convergence, will not depend on the 
shape of the particular distribution (they could also be different 
for different coefficients). 

It should be noticed that the values presented
in \efe{Degrassi}\ do not cover uncertainties associated to QCD
contribution in electroweak corrections. The dominant part of it is
included in $\delta_{QCD}$, the relevant correction in the electroweak
parameter $\Delta \rho$. The uncertainty in $\delta_{QCD}$
will reflect itself in a correlated way in the various theoretical
coefficients. 

To judge the effect of possible correlations
in the values given in \efe{Degrassi} 
we   use the  method outlined
at the end of Sect.~2. The more rigorous results derived with this procedure
are practically identical to those obtained using average
values and standard deviations of each coefficient. This is shown in
table \ref{tab:A1sum} and \ref{tab:ren_sch} where the comparison of the two
methods is presented. In the combined final $A_1$ result we report an 
additional digit to test the accuracy due to rounding. Also the final 
shape of the p.d.f. of $A_1$ obtainable from \equ{eq:fA1av}
is Gaussian with a good degree of approximation, since it is the average
of three Gaussians (each of which is 
justified by the central limit theorem) and
the closeness of their centers is much smaller than their widths.  
Given this situation, we  present our result as a function
of average coefficients and of their
standard deviations, because this method shows 
the sensitivity of $A_1$ to the various parameters in a clear way. 
\section{Results from the  indirect determination}\label{sec:res_dir}
The determinations of $M_W$  is presented in table \ref{tab:MW}.
The two values reported are obtained 
for  $(\Delta\alpha)_h^{EJ}$ 
 ($M_W = 80.375 \pm 0.027$ GeV) and 
$(\Delta\alpha)_h^{DH}$ 
 ($M_W = 80.366 \pm 0.025$ GeV). 
The two results are consistent and both 
are well in agreement with the experimental determinations given
in (\ref{eq:M_W_kin}) and (\ref{eq:M_W_NuTeV}). The uncertainty on the 
indirect $M_W$ 
determination is still a factor $\approx 2$ better 
than the present direct experimental result.
The table also shows  the contribution to the total 
uncertainty of each input quantity, with the sign of the derivative 
calculated in the reference point. This information allows 
the result to be corrected if any input quantity 
slightly  changes in expected value or standard deviation.
Through the entries in table \ref{tab:MW} we can 
estimate the  shift in the predicted central value due 
to unknown QCD effects in electroweak corrections. Indeed
a variation  in  $\delta_{QCD}$
introduces a shift on the calculated $s^2_{eff}$ and $M_W$ of
$\delta s^2_{eff}  \approx  
         - 1 \cdot 10^{-7}\, \delta (\delta_{QCD})\, M_t^2 $ and 
$ \delta M_W 
  \approx  2 \cdot 10^{-5}\, \delta (\delta_{QCD})\, M_t^2 $ (GeV).
For $M_t= 175$ GeV, $\delta(\delta_{QCD})$  has been estimated 
$\approx 5.2 \cdot 10^{-3}$ \cite{PhSi}. This induces
shifts in the  values of $s^2_{eff}$ and $M_W$ that amount
to $-1.8 \cdot 10^{-5}$ and $3.1$ MeV respectively. 
Using table \ref{tab:MW} one finds an  additional uncertainty
in the predicted $M_W$ central value, $\delta M_W \approx 1$ MeV.

The determination of $M_H$ from the effective mixing angle and $M_W$
separately is presented in table \ref{tab:Jeg} ($(\Delta\alpha)_h =
(\Delta\alpha)_h^{EJ}$) and table \ref{tab:A1sum} ($(\Delta\alpha)_h =
(\Delta\alpha)_h^{DH}$). All values are given in TeV to reduce the number 
of digits to the significant ones. The table also shows the 
combined determination. The values of
$A_1^s$ and $A_1^w$ are in agreement within uncertainty. 
However, the
$M_W$  determination is much less precise and the effect of combining it
with $A_1^s$ has almost a negligible impact on the determination
of the Higgs mass from $s^2_{eff}$, also because of the correlation between 
them. In case of slight variation of the central values of the input 
quantities the $A_1$ result   can be corrected using the
information provided in the tables. However the same procedure cannot be 
applied in case of changes of the standard deviations because $A_1$
is obtained through a combination where the inverse of variances enters.
For the same reason input quantities with large 
uncertainty are dumped in the combination and therefore they give
a small contribution to the total $A_1$ uncertainty.
In the various tables, combination results  are indicated by
``$\ast$''  below the relevant column. 

Among the various observables whose theoretical prediction depend 
upon $M_H$, given the present values of the $A_2$--$A_4$ quantities,
$s^2_{eff}$ is by far the most effective in constraining $M_H$.
Any other, like e.g.~$M_W$ or the leptonic width, has a very modest weight 
in a combined analysis. This fact  justifies our choice of considering
only one observable, $M_W$, in addition to $s^2_{eff}$.
This situation will not change in the 
near future. In fact,  a $W$ mass as effective in the $M_H$ indirect 
determination as the present $s^2_{eff}$  requires not only
a very precise $M_W$ result ($\sigma(M_W) \leq 25$ MeV) but also a reduction
in $M_t$ uncertainty ($\sigma(M_t) \leq 2.5$ GeV), as already pointed out
in \efe{DeZeu}. 
\begin{figure}[t]
\begin{center}
\epsfig{file=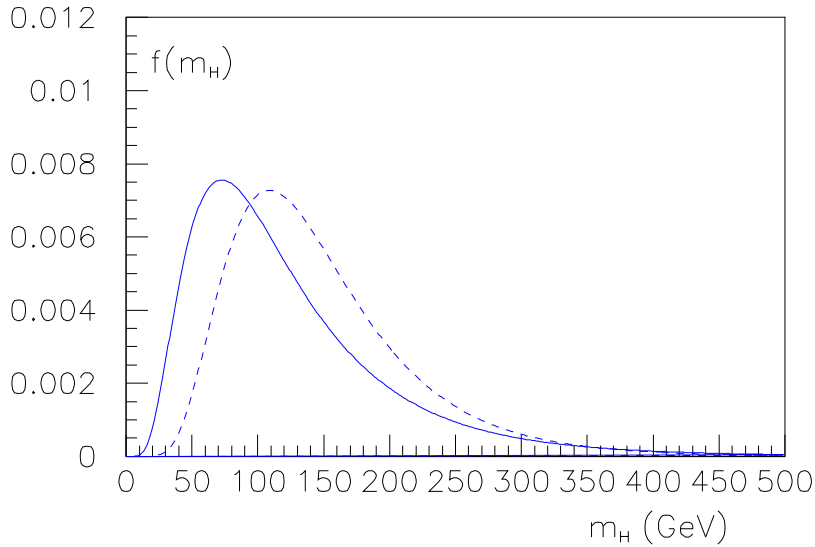,clip=,width=\linewidth} 
\end{center}
\caption{\sf Probability distribution function for the Higgs mass from
precision measurements. The solid line is obtained employing
$(\Delta\alpha)_h= 0.02804 \pm 0.00065$, the dashed one with 
$(\Delta\alpha)_h= 0.02763 \pm 0.00016$}
\label{fig:pdf_in}
\end{figure}

Figure \ref{fig:pdf_in} presents the p.d.f.~of $M_H$ obtained using only the 
indirect information. The comparison of the two curves 
shows that the use of a higher central value for $(\Delta\alpha)_h$ 
(i.e.~$(\Delta\alpha)_h^{EJ}$)\ tends to concentrate more the
probability towards smaller values of $M_H$. This can be understood
from the negative derivative of $M_H$ with respect to $(\Delta \alpha)_h$,
shown in tables \ref{tab:Jeg} and \ref{tab:A1sum}.
Indeed, in this case 
the median of the distribution is $M^{50}_H = 0.10$ TeV while the 
analysis performed employing $(\Delta\alpha)_h^{DH}$ gives for the same
quantity a result $\approx 0.3$ TeV higher, which is still less than
half a standard deviation of the distribution.
It is interesting to note that although the $A_1$ 
expected values and standard deviations in table \ref{tab:Jeg}  and 
\ref{tab:A1sum} are different, they actually give 
very close 95\,\% probability upper limit, $M_H^{95}$.
Similarly, while 
 $(\Delta\alpha)_h^{DH}$ 
has an uncertainty that is approximately
$4$ times smaller  than $(\Delta\alpha)_h^{EJ}$, 
the standard deviations
of the two $M_H$ p.d.f.'s are very close too.
Finally, we note that a variation in $\delta_{QCD}$ of the
order of magnitude  discussed above increases $M_H^{95}$ by
$\approx 10$ GeV.
\section{Results including the direct search}\label{sec:combinazione}
An extensive program to look  for 
evidence of Higgs production in $e^+e^-$ collision
has been pursued at LEP during the last decade.
Presently results for Higgs searches by all
four LEP collaborations are available  for
center of mass energies up to $\sqrt{s} = 183$ GeV 
\cite{ALEPH144,DELPHI95,L3052,OPAL173}.
The negative outcome of these searches has been 
reported as a combined 89.8 GeV ``$95\,\%\, C.L.$'' lower bound 
\cite{Vancouver,Dilodovico}.
Unfortunately this result has no simple probabilistic
interpretation regarding  the Higgs mass \cite{maxent98}. 
The operational definition of the limit  is expressed in terms
of a test-statistic, $X$, based on the number of selected events and their
distribution in a variable that discriminates signal from background, whose
value measured in the data, $X_{obs}$, is compared to that obtained on the
basis of a large number of ``simulated gedanken experiments'' \cite{LEP4}, or 
\be
CL_s (m_H) = \frac{CL_{s+b} (m_H)}{CL_{b} (m_H)}= 
\frac{P(X_{s+b} (m_H) \leq X_{obs})}{P(X_b (m_H) \leq X_{obs})}~~. 
\label{CLLep} 
\ee
The $C.L.$ for the signal + background  hypothesis, $CL_{s+b}$, 
is  defined as 
the probability that the test-statistic is less than or equal to $X_{obs}$,
 where the p.d.f of
$X_{s+b}$ is obtained by the Monte Carlo generation of experiments in which
a signal with mass $m_H$  is introduced in addition to the background. 
The corresponding $C.L.$ for the signal, $CL_{s}$, is then obtained 
by normalizing $CL_{s+b}$ to $CL_b$, the $C.L.$ for the background 
only hypothesis, where  the p.d.f.~of $X_b$ is obtained
similarly to that of $X_{s+b}$ but generating experiments with no signal.
One can realize that the $C.L.$ curves 
do not correspond to the likelihood 
of the observed data as a function of the Higgs  mass (i.e.
$f(\mbox{\it dir.}\,|\,m_H)$). Therefore, the published combined 
limit cannot be translated into a number 
which could be used in an unambiguous way in our analysis. 

As outlined in section \ref{sec:analysis}, the information from 
the direct search could easily be incorporated into the analysis 
if we had the likelihood or, more simply, 
the ${\cal R}$ function. We note here that the
test-statistic $X_{obs}$ of method A of \efe{LEP4}\ actually corresponds
to  ${\cal R}$. The publication of its value as a function of
$m_H$ would be sufficient for  a complete analysis to be performed. 
However, at the moment, this information is not available. 

Although we are not in a position to make a detailed analysis,
we can estimate the effect of the direct search results
on the final p.d.f.~of $m_H$ by employing the simplified  likelihood
given in \equ{eq:lik_Poisson} and the public  values of 
observed number of events, expected backgrounds and efficiencies.
With respect to this a few remarks are in order: 
{\it i)} in the function ${\cal R}$ the important effect is
always given by the data at the last energy point available. In fact,
because the final likelihood of searches at different energies is given
by the product of the individual likelihoods at the various $\sqrt{s}$, in 
the region of $M_H$ values close to $M_{K_{eff}}$ of the highest
energy point, the region which we are interested in, 
 only the corresponding ${\cal R} $ will be relevant
because those of lower energies are already saturated to 1. In our 
study we thus only use  data  at $\sqrt{s} = 183$ GeV 
\cite{ALEPH144,DELPHI95,L3052,OPAL173}.
{\it ii)} Differently from the other three experiments,
L3  uses  selection criteria that depend
on $m_H$. Correspondingly the number of selected events and
backgrounds will be functions of $m_H$. This situation is not compatible with 
our simplified likelihood. Therefore we do not consider the  L3 data.
{\it iii)} \equ{eq:lik_Poisson} does not contain any  information
related  to the distribution  of the signal and the backgrounds. At the
level at which we can perform our analysis using only public results, shape
effects cannot be taken into account. However,  the observed number of 
events in the various channels reported by the three collaborations 
are always either zero or few units. Given this situation  we expect that it 
is the event counting which gives most of the constraint. 
\begin{figure}[t]
\begin{center}
\epsfig{file=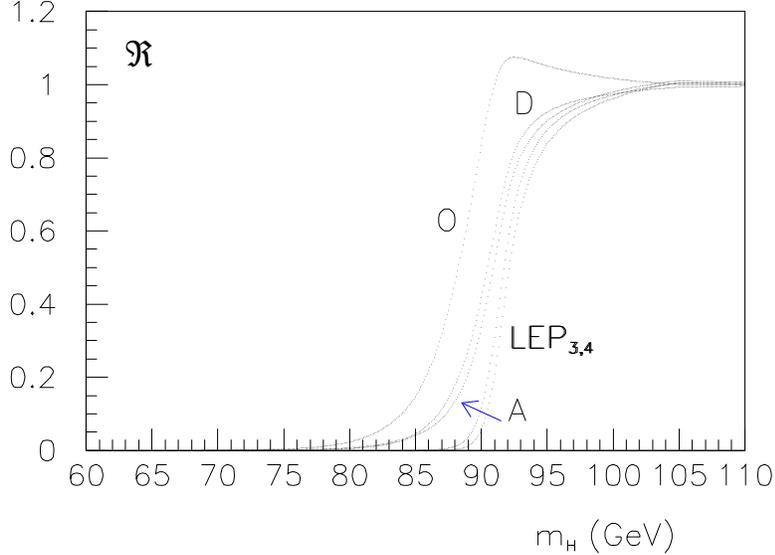,clip=,width=\linewidth} 
\end{center}
\caption{\sf ${\cal R}$ vs.~$m_H$ for the search at $\sqrt{s}= 183$ GeV.
The curves A, D, O correspond to the single experiment. The  
${\rm LEP}_{3}$ line represents the combination of the three.
The curve ${\rm LEP}_4$ is obtained by raising the ${\rm LEP}_3$ one
to the $4/3$ power.}
\label{fig:lik_lep4}
\end{figure}

Figure \ref{fig:lik_lep4} presents the ${\cal R}$ curves 
for the three experiments, ALEPH (A), DELPHI (D) and OPAL (O).
Each curve is obtained by multiplying the 
${\cal R}$'s of the individual search channels of the experiment. The 
overall ${\cal R}$ for the combination of three experiments is also
shown labeled as ${\rm LEP}_3$. As already said, we  cannot use L3
data. To try to take into account the L3 
contribution we make  the assumption 
that the L3 results are on  average similar to those of the other 
experiments.  We then roughly estimate the effect of L3, simply raising
the combined ${\cal R}$ of the other three experiments to the
$4/3$ power. The corresponding curve is presented in 
Fig.~\ref{fig:lik_lep4} marked ${\rm LEP}_4$. 
We note that the  OPAL ${\cal R}$
curve presents a bump  which is connected to a small excess of the  
observed number of events with respect to the expected background in 
the $q\bar{q}H$ channel. This bump is
not particularly significant for two reasons: i) it is not very pronounced
and therefore when the OPAL ${\cal R}$ is combined with 
$f(m_H\,|\,\mbox{\it ind.})$ the probability in the corresponding Higgs mass 
region is not particularly enhanced. 
ii) Our analysis is  based  on the event counting only and ignores
the distribution of the signal and background. When the latter information
is also taken into account it is not unlikely  that this bump 
disappears. 

According to \equ{eq:Bayes} the final p.d.f.~is obtained by  combining 
the p.d.f.~coming from precision measurements 
(Fig.~\ref{fig:pdf_in})
with the likelihood derived from  the LEP data (Fig.~\ref{fig:lik_lep4}), 
rescaled to the function ${\cal R}$, or
\be 
f(m_H\,|\,\mbox{\it dir.} \,\&\,\mbox{\it ind.}) =
\frac{ {\cal R}(m_H)\, f(m_H\,|\,\mbox{\it ind.})}
    {\int_{0}^\infty {\cal R}(m_H)\, f(m_H\,|\,\mbox{\it ind.})\, dm_H}~.
\label{fine}
\ee
\begin{figure}
\begin{center}
\epsfig{file=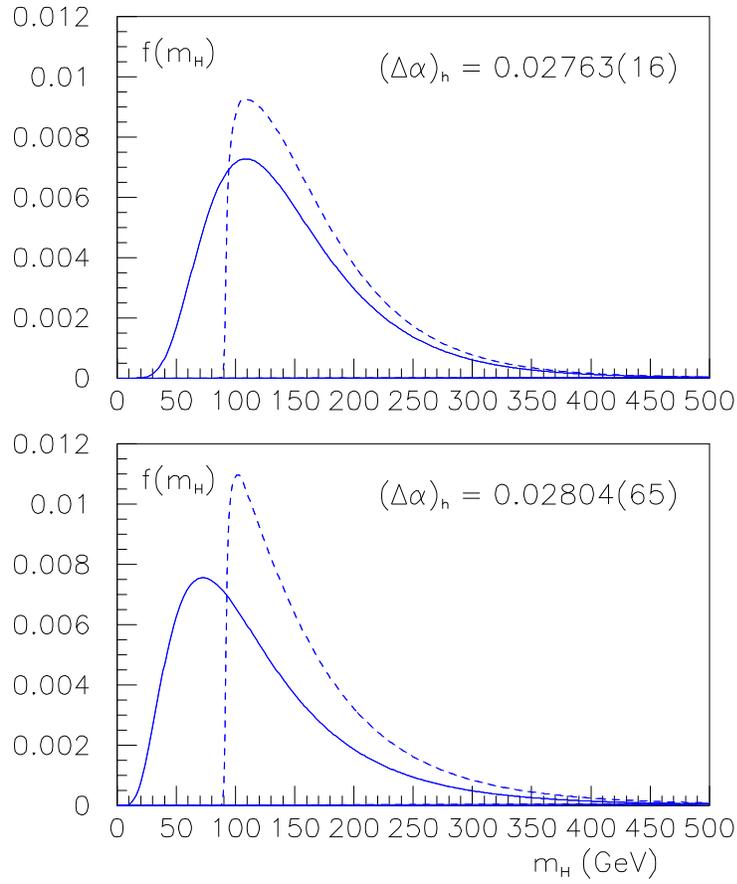,clip=,width=0.9\linewidth} 
\end{center}
\caption{\sf Probability distribution functions using only indirect information
(solid line) and employing also the experimental results from direct searches
(dashed one). }
%a) $(\Delta \alpha)_h = 0.02804 \pm 0.00065$; b)
%$(\Delta \alpha)_h = 0.02763 \pm 0.00016$.}
\label{fig:pdf_fin}
\end{figure}
The result is shown in figure \ref{fig:pdf_fin} where 
$f(m_H\,|\,\mbox{\it dir.} \,\&\,\mbox{\it ind.})$ 
is compared to $f(m_H\,|\,\mbox{\it ind.})$
for each assumption on $(\Delta \alpha)_h$.
To judge the sensitivity of 
$f(m_H\,|\,\mbox{\it dir.} \,\&\,\mbox{\it ind.})$  to  
the likelihood   we have actually evaluated \equ{fine} 
using  two different ${\cal R}$'s (${\rm LEP}_3$  and ${\rm LEP}_4$). 
The two resulting curves are indistinguishable as could be expected given
that the two ${\cal R}$'s differ practically by a shift of 
$\approx 1/2$ GeV.
To envisage a more different case,  we  compare
the final distribution presented in fig.~\ref{fig:pdf_fin} with that 
we obtain in our simplified analysis using only OPAL data. 
The difference in the expected value amounts to
4-5 GeV, depending on the value of $(\Delta \alpha)_h$ chosen,
 while standard deviations are the same. The
``OPAL'' $M_H^{95}$ is 3-6 GeV lower than the one reported in table 
\ref{tab:results} where our final results are presented. The closeness
of the various quantities shows 
that our results do not depend critically upon the details of 
the analysis.

As expected, the inclusion  of the direct search information in the
Higgs mass probability analysis drifts the p.d.f.~towards higher
values of $M_H$, changing its shape such that the probability of
$M_H$ values below 95 GeV drops to $\approx 3 \, \%$.
Table \ref{tab:results} 
summarizes our analysis  in terms of various convenient 
parameters of the distribution. Expected value, standard deviation, 
mode and median are not very sensitive to the values of the hadronic 
contribution to the vacuum polarization. Also in both cases, $75\, \%$ of
the probability is concentrated in the region $M_H < 0.20$ TeV.
Instead the choice of $(\Delta \alpha)_h$ affects the tail of the
distribution with $(\Delta \alpha)_h^{EJ}$ producing a longer one.
Indeed while the use of only indirect information gives  same 
values of $M_H^{95}$, when the direct one is also included, the  $M_H^{95}$
obtained using $(\Delta \alpha)_h^{EJ}$ is $\sim 0.3$ TeV higher
than the corresponding number for $(\Delta \alpha)_h^{DH}$, and this
effect is even more pronounced for $M_H^{99}$.
%%%%%%%%%%%%%%%%%%%%%%%%%%%%%%%%%%%%%
%%%%%%%%%%%%%%%%%%%%%%%%%%%%%%%%%%%%%
%%%%%%%%%%%%%%%%%%%%%%%%%%%%%%%%%%%%%
%%%%%%%%%%%%%%%%%%%%%%%%%%%%%%%%%%%%%
\section{Conclusions}
We have presented a method that allows the Higgs mass to be constrained 
combining
together the indirect information coming from precision measurements and
accurate calculations, with the results of  the search experiments
currently being carried out at LEP. The method makes use of the Bayes' theorem
which allows  the p.d.f.~obtained from precision measurements 
to be augmented with the direct
experimental information through the likelihoods of the various search
channels. Such likelihoods should be provided by
the experiments, possibly in the form of the ${\cal R}$ function that
is very convenient for comparing and combining the various informations
and it has also an intuitive interpretation because of its limit to 
the step function of the ideal case. 
Using the simplified form  of  \equ{eq:lik_Poisson} and the public results
concerning   observed number of events, backgrounds and
efficiencies of the ALEPH, DELPHI, OPAL experiments for $\sqrt{s} = 183$ GeV,
we  have  derived a p.d.f.~for the Higgs boson mass that includes the 
direct search constraint. 
Several parameters of the distribution has been reported and
we have  verified that the main conclusions do not 
depend to any great degree upon the detailed form of the likelihood.
 
It should be clearly stated that our results are derived
under the assumption of the validity of the SM and using as input quantities
those described in section \ref{sec:input}. Our analysis  does not apply
to  different frameworks, like
for example the Minimal Supersymmetric Standard Model except the case
when all SUSY particles are decoupled. Concerning the input quantities,
our results  depend largely upon the 
experimental value for $s^2_{eff}$ we have taken (\equ{seff}). It seems to
us reasonable to employ the combined LEP+SLD value although, given the less
than perfect agreement between the two most precise
determinations, the suspicion that some 
measurements are affected by not yet understood systematic errors exists. 

The last run of the LEP machine was performed at $\sqrt{s} = 189$ GeV.
Results on Higgs searches at this energy are still preliminary and only
DELPHI has presented a somewhat detailed analysis of the searches in the 
$H+Z^\circ$ channel \cite{LEPnov}. In the case of negative results by the 
other collaborations similar to those reported by
DELPHI, we can make a rough estimation of the output of the search at
$\sqrt{s}= 189$ GeV by saying that the ${\cal R}$ will move  5-6 GeV
towards higher values of $M_H$. The final 
$f(m_H\, |\, \mbox{\it dir.}\, \&\, \mbox{\it ind.} )$ will be
correspondingly shifted in the same direction. 
\vspace{2cm} \\

We wish to thank F.~Di Lodovico, R.~Faccini and G.~Ganis for useful 
communications and discussions.

\begin{table}
\begin{center}
\begin{tabular}{|c|ll|ll|}\hline
& & & &   \\
$X_i$&E[$X_i$]&$\sigma(X_i)$&
\multicolumn{2}{|c|}
{$\frac{\partial M_W}{\partial X_i}\cdot 
\sigma(X_i)$}  \\ 
& & & \multicolumn{2}{|c|}{(MeV)} \\\hline
& & & &   \\
& & & \multicolumn{1}{|c}{$(\Delta\alpha)_h^{EJ}$} & 
                         \multicolumn{1}{c|}{$(\Delta\alpha)_h^{DH}$} \\
$(s^2_{eff})_\circ$ & 0.231525& 0.000015 & $+1.7$  & $+1.8$\\
$c_1$      & (5.23    & 0.04)$\cdot 10^{-4}$   &  {\small$+1.8\cdot 10^{-3}$}
                                         & $+0.13$      \\
$c_2$      & (9.860   & 0.003)$\cdot 10^{-3}$    & {\small$+4.7\cdot 10^{-4}$}
                                         & {\small$-4.7\cdot 10^{-3}$}\\
$c_3$      & (2.74    & 0.06) $\cdot 10^{-3}$    & {\small$+6.1\cdot 10^{-2}$}
                                         & {\small$+6.5\cdot 10^{-2}$}\\
$c_4$      & (4.47    & 0.06) $\cdot 10^{-4}$    & {\small$+5.6\cdot 10^{-3}$}
                                         & {\small$+6.1\cdot 10^{-3}$} \\
& & & &   \\
$M^\circ_W$/GeV     & 80.3813 & 0.0012   & $+1.2$ & $+1.2$ \\
$d_1$               & 0.0578  & 0.0004         & {\small$-1.6\cdot 10^{-3}$}
                                         & $-0.11$   \\
$d_2$               & 0.5177  & 0.0006          & {\small$-8.6\cdot 10^{-4}$}
                                         & {\small$+7.9\cdot 10^{-3}$}  \\
$d_3$               & 0.540   & 0.003           & {\small$-2.7\cdot 10^{-2}$}
                                         & {\small$-2.7\cdot 10^{-2}$}\\
$d_4$               & 0.0850  & 0.0003          & {\small$-2.5\cdot 10^{-3}$}
                                         & {\small$-2.5\cdot 10^{-3}$}\\
$d_5$               & 0.00793 & 0.00012         & {\small$-2.0\cdot 10^{-6}$}
                                         & {\small$-9.4\cdot 10^{-3}$}\\
& & & &  \\
$(\Delta\alpha)_h$       & 0.02804& 0.00065& $+13$&\\
$(\Delta\alpha)_h$       & 0.02763& 0.00016& & $+3.7$\\
& & & &  \\
$M_t$/GeV                     & 174.2   & 4.8      & $+13$  & $+12$\\
$\alpha_s(M_Z)$               & 0.119   & 0.002    & $-0.60$& $-0.54$ \\
$s^2_{eff}$                   & 0.23157 & 0.00018  & $-20$  & $-21$ \\
& & & &  \\
\hline
& & & &  \\
$M_W$/GeV          & {\bf 80.375} & {\bf 0.027} 
                                         & 
\multicolumn{1}{|c}{$\leftarrow$}    &            \\
& & & &   \\
$M_W$/GeV          & {\bf 80.366} & {\bf 0.025} 
                                         & & 
\multicolumn{1}{c|}{$\leftarrow$}              \\
& & & &  \\
\hline  
\end{tabular}
%} %fine di \small
\end{center}
\caption{\sf $M_W$ determination.}
\label{tab:MW}
\end{table}

\begin{table}
\begin{center}
\begin{tabular}{|c|ll|lll|}\hline
& & & & &  \\
$X_i$&E[$X_i$]&$\sigma(X_i)$&
\multicolumn{3}{|c|}
{$\frac{\partial A_1}{\partial X_i}\cdot 
\sigma(X_i)$}  \\ 
& & & & &  \\
\hline
& & & & &  \\
 &&&\multicolumn{1}{|c}{``$s^2_{eff}$''} & \multicolumn{1}{c}{``$M_W$''}&
\multicolumn{1}{c|}{Comb.} \\ 
& & & & &  \\
$(s^2_{eff})_\circ$ & 0.231525& 0.000015 & $-0.029$          &    
                                         & $-0.024$    \\
$c_1$               & (5.23    &   0.04)$\cdot 10^{-4}$     
                               & {\small$-3.1\cdot 10^{-5}$}&   
                               & {\small$-1.0\cdot 10^{-3}$} \\
$c_2$               & (9.860   &  0.003)$\cdot 10^{-3}$
                               & {\small$-8.2\cdot 10^{-6}$}&
                               & {\small$+1.3\cdot 10^{-5}$} \\
$c_3$               & (2.74    &   0.06)$\cdot 10^{-3}$     
                               & {\small$-1.0\cdot 10^{-3}$}&
                               & {\small$-6.3\cdot 10^{-4}$} \\
$c_4$               & (4.47    & 0.06)$\cdot 10^{-4}$     
                               & {\small$-9.7\cdot 10^{-5}$}&
                               & {\small$-8.2\cdot 10^{-5}$} \\
& & & & &  \\
$M^\circ_W$/GeV   & 80.3813 & 0.0012   &          & $+0.021$ 
                                         & {\small$+4.0\cdot 10^{-3}$}  \\
$d_1$             & 0.0578    & 0.0004   &        &{\small$-2.0\cdot 10^{-3}$}
                                         & {\small$+5.4\cdot 10^{-4}$} \\
$d_2$             & 0.5177    & 0.0006   &        &{\small$-1.5\cdot 10^{-5}$}
                                         & {\small$-2.6\cdot 10^{-5}$}\\
$d_3$             & 0.540     & 0.003    &        &{\small$-4.8\cdot 10^{-4}$}
                                         & {\small$-3.5\cdot 10^{-4}$}\\
$d_4$             & 0.0850    & 0.0003   &        &{\small$-4.5\cdot 10^{-5}$}
                                         & {\small$-8.9\cdot 10^{-6}$}\\
$d_5$             & 0.00793   & 0.00012  &        &{\small$-1.7\cdot 10^{-4}$}
                                         & {\small$+1.2\cdot 10^{-4}$}\\
& & & & & \\
$(\Delta\alpha)_h$  & 0.02804 & 0.00065  & $-0.44 $ & $-0.21$  
                                         & $-0.41 $ \\
$M_t$/GeV           & 174.2   & 4.8      & $+0.29$  & $+0.47$ 
                                         & $+0.33$  \\
$\alpha_s(M_Z)$     & 0.119   & 0.002    & $-0.014$ & $-0.026$ 
                                         & $-0.017$ \\
$s^2_{eff}$         & 0.23157 & 0.00018  & $+0.34$  &       
                                         & $+0.29$   \\
$M_W^k$/GeV         & 80.39   & 0.06     &          & $-0.82$ 
                                         & $-0.15$ \\
$M_W^\nu$/GeV       & 80.25   & 0.11     &          & $-0.45$ 
                                         & $-0.080$ \\
 & & & &\multicolumn{1}{c}{{\Large $\ast$}} 
       &\multicolumn{1}{c|}{{\Large $\ast$}} \\
\hline
& & & & & \\
$A^s_1$            & $0.00$     & $\rightarrow$
                                         & 0.63 &   &           \\
$A_1^w$            & $+0.28$       & $\rightarrow$
                                         &       & 1.07 &        \\
& & &
\multicolumn{2}{|c}
{$\underbrace{\hspace{3.5cm}}_{\rho=+0.34}$}  & \\
\hline
& & & & & \\
$A_1  $    & {\bf 0.05}& $\rightarrow$   &
\multicolumn{2}{|c}{{\bf 0.61}} & \multicolumn{1}{c|}{$\leftarrow$}           \\
& & & &  &\\ \hline
& & & &  &\\
$M_H$/TeV    & {\bf 0.13}     & {\bf 0.08} &  
\multicolumn{3}{|c|}{$\left(
\hat{M}_H = 0.07\,\mbox{TeV},\      
M^{50}_H= 0.10\,\mbox{TeV}  
\right)$}              \\
& & & &  &\\
\hline  
\end{tabular}
%} %fine di \small
\end{center}
\caption{\sf Summary of indirect information. 
$A_1 \equiv \ln(M_H/100\,\mbox{GeV})$,
$(\Delta\alpha)_h=(\Delta\alpha)_h^{EJ}$. See text for the meaning of 
``$\ast$''. }
\label{tab:Jeg}
\end{table}

\begin{table}
\begin{center}
\begin{tabular}{|c|ll|lll|}\hline
& & & & &  \\
$X_i$&E[$X_i$]&$\sigma(X_i)$&
\multicolumn{3}{|c|}
{$\frac{\partial A_1}{\partial X_i}\cdot 
\sigma(X_i)$}  \\ 
& & & & &  \\
\hline
& & & & &  \\
 &&&\multicolumn{1}{|c}{``$s^2_{eff}$''} & \multicolumn{1}{c}{``$M_W$''}&
\multicolumn{1}{c|}{Comb.} \\ 
& & & & &  \\
$(s^2_{eff})_\circ$ & 0.231525& 0.000015 & $-0.029$          &    
                                         & $-0.026$    \\
$c_1$               & (5.23    &   0.04)$\cdot 10^{-4}$     
                               & {\small$-2.1\cdot 10^{-3}$}&   
                               & {\small$-2.3\cdot 10^{-3}$} \\
$c_2$               & (9.860   &  0.003)$\cdot 10^{-3}$
                               & {\small$+7.6\cdot 10^{-5}$}&
                               & {\small$+7.0\cdot 10^{-5}$} \\
$c_3$               & (2.74    &   0.06)$\cdot 10^{-3}$     
                               & {\small$-1.0\cdot 10^{-3}$}&
                               & {\small$-8.4\cdot 10^{-4}$} \\
$c_4$               & (4.47    & 0.06)$\cdot 10^{-4}$     
                               & {\small$-9.7\cdot 10^{-5}$}&
                               & {\small$-8.9\cdot 10^{-5}$} \\
& & & & &  \\
$M^\circ_W$/GeV   & 80.3813 & 0.0012   &          & $+0.020$ 
                                         & {\small$+1.8\cdot 10^{-3}$}  \\
$d_1$             & 0.0578    & 0.0004   &        &{\small$-2.8\cdot 10^{-3}$}
                                         & {\small$+5.0\cdot 10^{-6}$} \\
$d_2$             & 0.5177    & 0.0006   &        &{\small$+1.4\cdot 10^{-4}$}
                                         & {\small$+1.1\cdot 10^{-5}$}\\
$d_3$             & 0.540     & 0.003    &        &{\small$-4.7\cdot 10^{-4}$}
                                         & {\small$-1.6\cdot 10^{-4}$}\\
$d_4$             & 0.0850    & 0.0003   &        &{\small$-4.3\cdot 10^{-5}$}
                                         & {\small$-4.1\cdot 10^{-6}$}\\
$d_5$             & 0.00793   & 0.00012  &        &{\small$-3.5\cdot 10^{-4}$}
                                         & {\small$+3.2\cdot 10^{-5}$}\\
& & & & & \\
$(\Delta\alpha)_h$  & 0.02763 & 0.00016  & $-0.11 $ & $-0.050$  
                                         & $-0.10 $ \\
$M_t$/GeV           & 174.2   & 4.8      & $+0.29$  & $+0.46$ 
                                         & $+0.30$  \\
$\alpha_s(M_Z)$     & 0.119   & 0.002    & $-0.014$ & $-0.025$ 
                                         & $-0.016$ \\
$s^2_{eff}$         & 0.23157 & 0.00018  & $+0.34$  &       
                                         & $+0.32$   \\
$M_W^k$/GeV         & 80.39   & 0.06     &          & $-0.79$ 
                                         & $-0.071$ \\
$M_W^\nu$/GeV       & 80.25   & 0.11     &          & $-0.43$ 
                                         & $-0.039$ \\
 & & & &\multicolumn{1}{c}{{\Large $\ast$}} 
       &\multicolumn{1}{c|}{{\Large $\ast$}} \\
\hline
& & & & & \\
$A^s_1$            & 0.28     & $\rightarrow$
                                         & 0.46 &   &           \\
$A_1^w$            & 0.41 & $\rightarrow$
                                         &       & 1.01 &        \\
& & &
\multicolumn{2}{|c}
{$\underbrace{\hspace{3.5cm}}_{\rho=+0.29}$}  & \\
\hline
& & & & & \\
$A_1  $    & $\mathbf{ 0.29_{1}}$& $\rightarrow$   &
\multicolumn{2}{|c}{$\mathbf{ 0.45_5}$} & \multicolumn{1}{c|}{$\leftarrow$} \\
& & & &  &\\ \hline
& & & &  &\\
$M_H$/TeV    & {\bf 0.15}     & {\bf 0.07} &
\multicolumn{3}{|c|}{$\left(
\hat{M}_H = 0.11\,\mbox{TeV},\     
M^{50}_H= 0.13\,\mbox{TeV} 
\right)$}              \\
& & & &  &\\
\hline  
\end{tabular}
\end{center}
\caption{\sf Like table \ref{tab:Jeg} but for 
$(\Delta\alpha)_h= (\Delta\alpha)_h^{DH}$. }
\label{tab:A1sum}
\end{table}

\begin{table}
\begin{center}
\begin{tabular}{|c|ll|lll|}\hline
& & & & &  \\
$X_i$&E($X_i$)&$\sigma(X_i)$&
\multicolumn{3}{|c|}
%{$\mbox{sign}\left[\frac{\partial A_1}{\partial X_i}\right]\cdot 
%\sigma_i(A_1)$}  \\ \hline
{$\frac{\partial A_1}{\partial X_i}\cdot 
\sigma(X_i)$}  \\ 
& & & & &  \\
\hline
& & & & &  \\
 &&&\multicolumn{1}{|c}{``$\overline{\mbox{MS}}$''} & 
\multicolumn{1}{c}{``OSI''}&
\multicolumn{1}{c|}{``OSII''} \\ 
& & & & &  \\
$(\Delta\alpha)_h$  & 0.02763 & 0.00016  & $-0.10 $ & $-0.10$  
                                         & $-0.10 $ \\
$M_t$/GeV           & 174.2   & 4.8      & $+0.31$  & $+0.31$ 
                                         & $+0.30$  \\
$\alpha_s(M_Z)$     & 0.119   & 0.002    & $-0.016$ & $-0.016$ 
                                         & $-0.015$ \\
$s^2_{eff}$         & 0.23157 & 0.00018  & $+0.32$  & $+0.32$      
                                         & $+0.32$   \\
$M_W^k$/GeV         & 80.39   & 0.06     & $-0.072$ & $-0.070$ 
                                         & $-0.070$ \\
$M_W^\nu$/GeV       & 80.25   & 0.11     & $-0.039$ & $-0.038$ 
                                         & $-0.038$ \\
 & &  &\multicolumn{1}{c}{{\Large $\ast$}} 
      &\multicolumn{1}{c}{{\Large $\ast$}}
      &\multicolumn{1}{c|}{{\Large $\ast$}} \\
%& & & & & \\
\hline
& & & & & \\
$A_1\,|\,R_1$ ($\overline{\mbox{MS}}$) & 0.318     & $\rightarrow$
                                         & 0.456 &   &            \\
$A_1\,|\,R_2$ (OSI)           & 0.293       & $\rightarrow$
                                         &       & 0.460 &        \\
$A_1\,|\,R_3$ (OSII)          & 0.262       & $\rightarrow$
                                         &       &       & 0.448  \\
& & & \multicolumn{3}{|c|}
{$\underbrace{\hspace{5.0cm}}_{\sigma^2(A_1)
=\frac{1}{3}\sum_i\sigma^2(A_1\,|\,R_i)+\sigma_E^2}$}   \\
& & & & & \\
\hline
& & & & & \\
$A_1$ (average)       & $\mathbf{ 0.29_1}$      & $\rightarrow$ &
\multicolumn{3}{|c|}{$\mathbf{0.45_5}$} \\
& & & & & \\
\hline  
\end{tabular}
\end{center}
\caption{\sf Renormalization scheme dependence. The results shown here
should be compared with those of table \ref{tab:A1sum}.}
\label{tab:ren_sch}
\end{table}

\begin{table}
\begin{center}
\begin{tabular}{|l|cc|cc|}\hline
& & & &   \\
  & \multicolumn{2}{|c|}{$(\Delta\alpha)_h= 0.02804(65)$} 
  & \multicolumn{2}{|c|}{$(\Delta\alpha)_h= 0.02763(16)$} \\
& & & &   \\ \hline
& & & &   \\
& $ (ind.)$ 
& $\left(\begin{array}{c} ind. \\ + \\ dir. \end{array}\right)$ 
&$(ind.)$ 
& $\left(\begin{array}{c} ind. \\ + \\ dir. \end{array}\right)$\\
& & & &   \\
$\mbox{\bf E}[\mathbf{M_H}]$/TeV  &  0.13   & {\bf 0.17}  &  0.15 &{\bf 0.17}\\
$\mathbf{\sigma(M_H)}$/TeV        &  0.08   & {\bf 0.08}  &  0.07 &{\bf 0.07}\\
$\hat{M}_H$/TeV                   &  0.07   & 0.10        &  0.11 &0.11\\
$\mathbf{M^{50}_H}$/TeV       &  0.10   & {\bf 0.15}  &  0.13 &{\bf 0.15}\\
& & & &   \\
$P(M_H\le 95\,\mbox{GeV})$          &44\,\% & 3.2\,\% & 23\,\% & 2.3\,\%  \\
$P(M_H\le 0.11\,\mbox{TeV})$          &53\,\% & 19\,\%    & 33\,\% & 16\,\%  \\
$P(M_H\le 0.13\,\mbox{TeV})$          &64\,\% & 38\,\%    & 48\,\% & 34\,\%  \\
$P(M_H\le 0.20\,\mbox{TeV})$          &86\,\% & 75\,\%    & 81\,\% & 76\,\%  \\
& & & &   \\
$\mathbf{M^{95}_H}$/TeV; $P(M_H\le M^{95}_H)\approx 0.95$ 
                                          &0.28 &{\bf 0.33}& 0.28& {\bf 0.30}\\
$\mathbf{M^{99}_H}$/TeV; $P(M_H\le M^{99}_H)\approx 0.99$ 
                                          &0.43 &{\bf 0.48}& 0.39& {\bf 0.40}\\
& & & &   \\
$\left\{\!\!\begin{array}{l} M_1/\mbox{TeV};\ P(M_H < M_1)\approx 0.16 \\
                 M_2/\mbox{TeV};\ P(M_H > M_2)\approx 0.16 \end{array}\right.$ 
& $\left\{\!\!\begin{array}{l} 0.06 \\ 0.19 \end{array}\right.$ 
& $\left\{\!\!\begin{array}{l} 0.11 \\ 0.23 \end{array}\right.$ 
& $\left\{\!\!\begin{array}{l} 0.08 \\ 0.21 \end{array}\right.$ 
& $\left\{\!\!\begin{array}{l} 0.11 \\ 0.23 \end{array}\right.$ \\
& & & &  \\
$\left\{\!\!\begin{array}{l} \mathbf{M_3}/\mbox{TeV};\ P(M_H < M_3)
        \approx 0.25 \\
    \mathbf{M_4}/\mbox{TeV};\ P(M_H > M_4)\approx 0.25 \end{array}\right.$ 
& $\left\{\!\!\begin{array}{l} 0.07\\ 0.16 \end{array}\right.$ 
& $\left\{\!\!\begin{array}{l} \mbox{\bf 0.12}\\ \mbox{\bf 0.20} 
              \end{array}\right.$ 
& $\left\{\!\!\begin{array}{l} 0.10\\ 0.18 \end{array}\right.$ 
& $\left\{\!\!\begin{array}{l} \mbox{\bf 0.12}\\ \mbox{\bf 0.20} 
              \end{array}\right.$ \\
& & & &  \\
\hline  
\end{tabular}
\end{center}
\caption{\sf Summary of the direct plus indirect information.}
\label{tab:results}
\end{table}
\end{document}